\documentclass[11pt,twoside]{article}

\usepackage{asp2006}
\usepackage{epsf}
\usepackage{graphicx}
\usepackage{lscape}

\begin{document}
\title{The hot and the cool otbursts in the symbiotic system \textit{AG Draconis}}
\author{M. Cika\l{}a$^1$, M. Miko\l{}ajewski$^1$, J. Osiwa\l{}a$^1$, T. Tomov$^1$, L. Leedj$\ddot{\mathrm{a}}$rv$^2$, M. Burmeister$^{2,3}$}
\affil{$^1$Centre for Astronomy of Nicolaus Copernicus University, Toru\'n, Poland}
\affil{$^2$Tartu Observatory, Estonia}
\affil{$^3$Department of Physic, University of Tartu}

\begin{abstract}
In this papper we present the analyses of the six (1998, 1997, 2001, 2002, 2003 and 2005) last outbursts of AG Draconis on the basis of low resolution visual spectroscopy. A new method to determine the Zanstra's temperature of the hot ionizing source from the optical H$\beta$ and HeII$\lambda4686$\AA\, emission lines has been used. As a results we obtained the evolution of the individual outburst on the H-R diagram.
\end{abstract}

\section{Background}

AG Dra is the best studied case of so-called ''yellow'' symbiotic binaries containing a hot white dwarf and a metal-poor K giant. The giant is significantly brighter than similar normal giants with the same temperature belonging to the disc population \citep{mik95}. It is a relatively active system among the other symbiotics. Its quiescence stages are almost regularly disrupted (in 1936, 1951, 1966, 1980, and 1994) with intervals of about 15 years series of several (2-5) outbursts \citet{bastian}. The last active phase consist of five outbursts between 1994 and 1998. In 2001 an unexpected series of three outbursts have been observed. New active stage has started in 2006, and was preceded by smaller outburst in 2005. Basing on the analysis of the 1980-86 series of outbursts \citet{gr} distinguished ''hot'' and ''cool'' outbursts.

The most of the spectroscopic and photometric observations were presented in \citet{leed}. A new set of observations was also obtained. Low resolution and objective prism spectroscopy, was carried out in Torun Observatory using Canadian Copernicus Spectrograph, and the flint prism, attached to the 60/90cm Schmidt-Cassegrain telescope. All the Torun spectra were converted to absolute flux scale using standard stars and available photometry. Additionally the equivalent widths published in \citep{leed} were converted to fluxes. The observed $H\beta$ fluxes were corrected for the self-absorption (in the HII regions), and the absorption (in the HI regions) effects.

\begin{figure}[!t]
\centering
\includegraphics[width=0.35\textwidth]{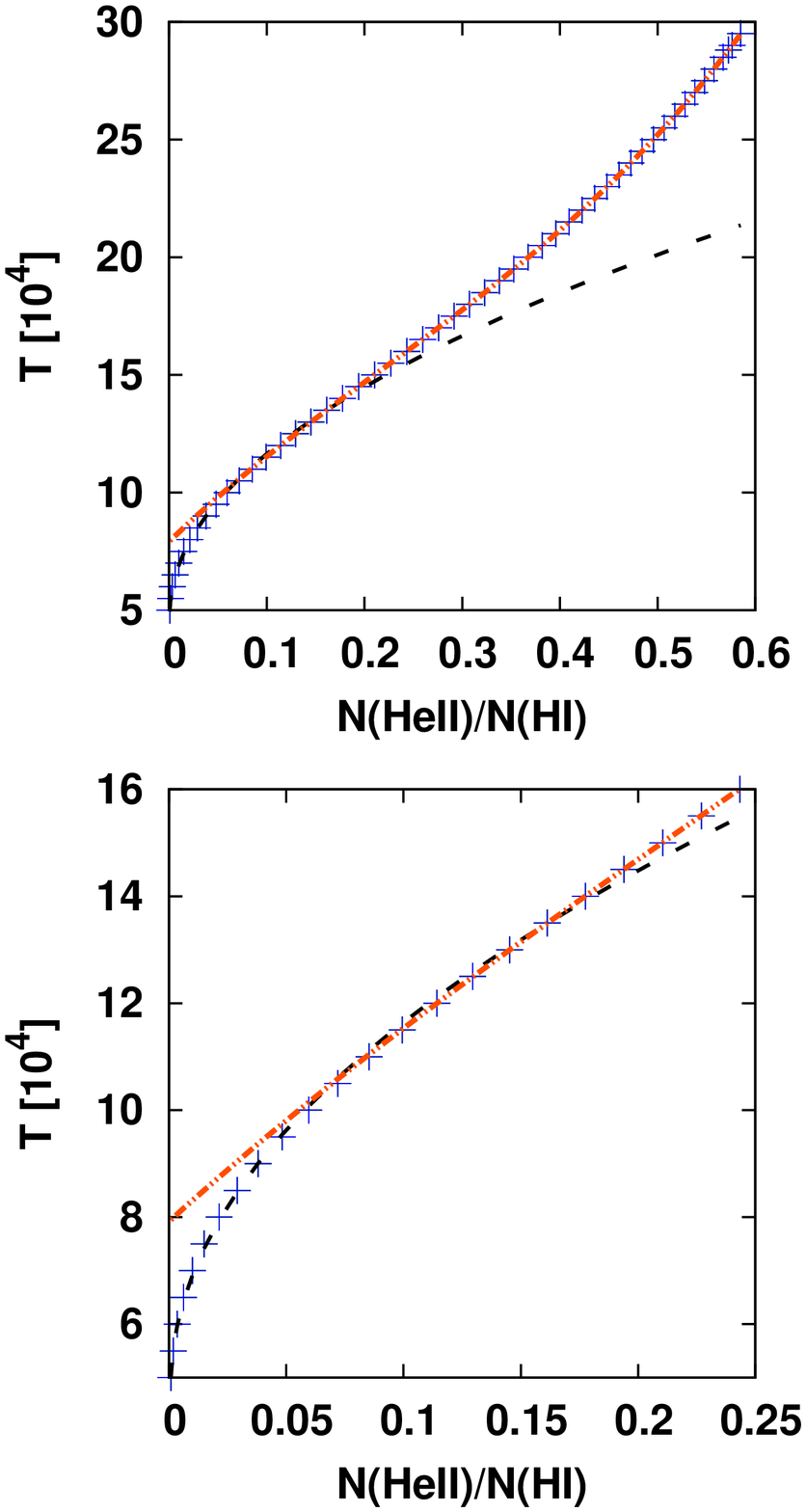}
\includegraphics[width=0.45\textwidth]{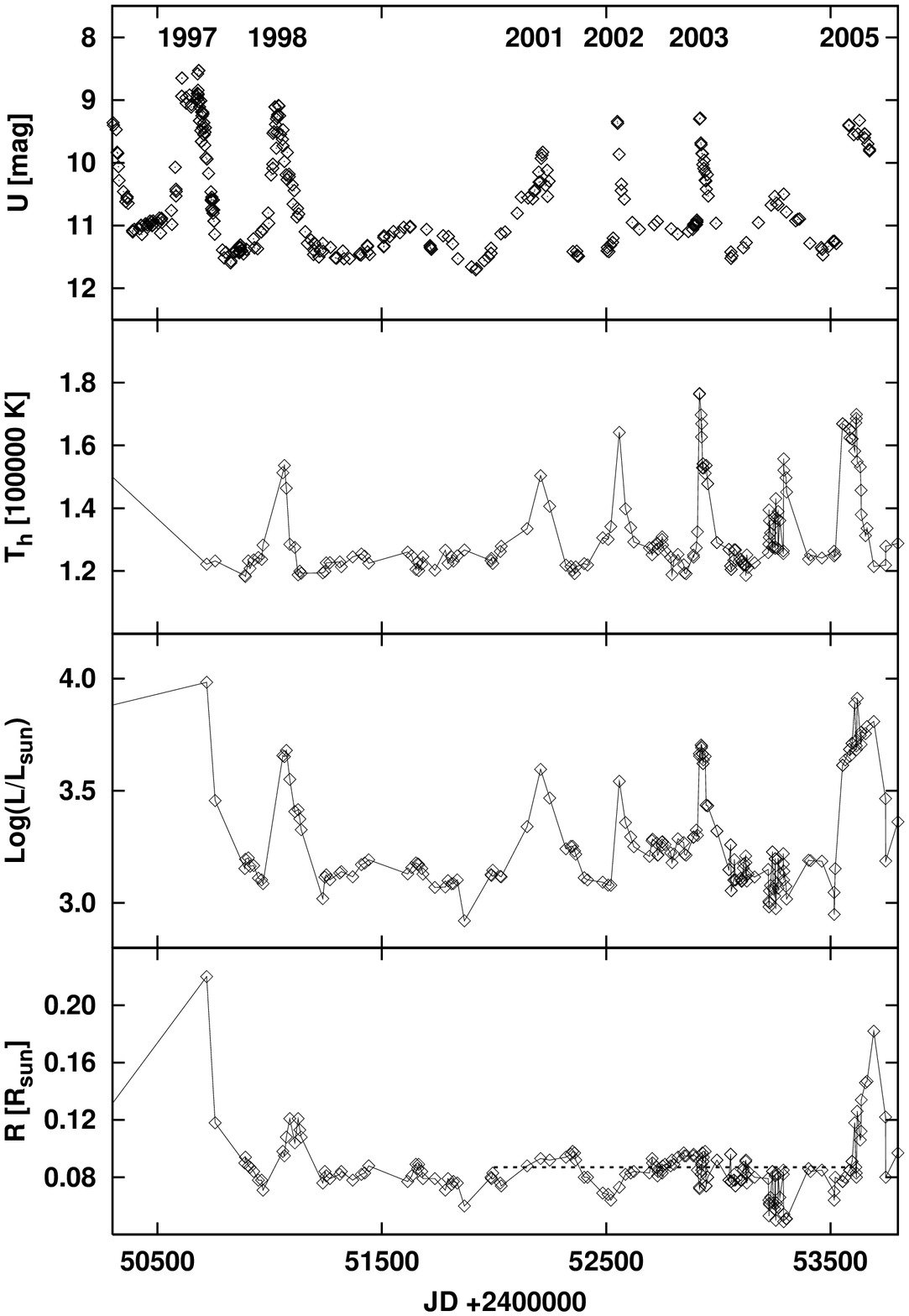}
\caption{Left upper panel: the ratio of HeII and HI number of emitted photons vs. the temperature for case \textit{B} recombination. The photon ratios calculated with Planck formula are shown with plus signs. The solid line represents the solution for temperature $50000 < T[K] < 140000$, the dashed line -- the solution for $100000 < T[K] < 300000$. Left bottom panel shows the details of low-range part of left upper picture. Right panel: The time evolution of the V-band photometry and the hot star parameters. Starting from upper panel we have V-band photometry of the system, the temperature, the luminosity and the radius of hot star. For details see in text.}
\end{figure} 


\section{Method}

\begin{figure}[!th]
   \centering
   \includegraphics[width=0.7\textwidth]{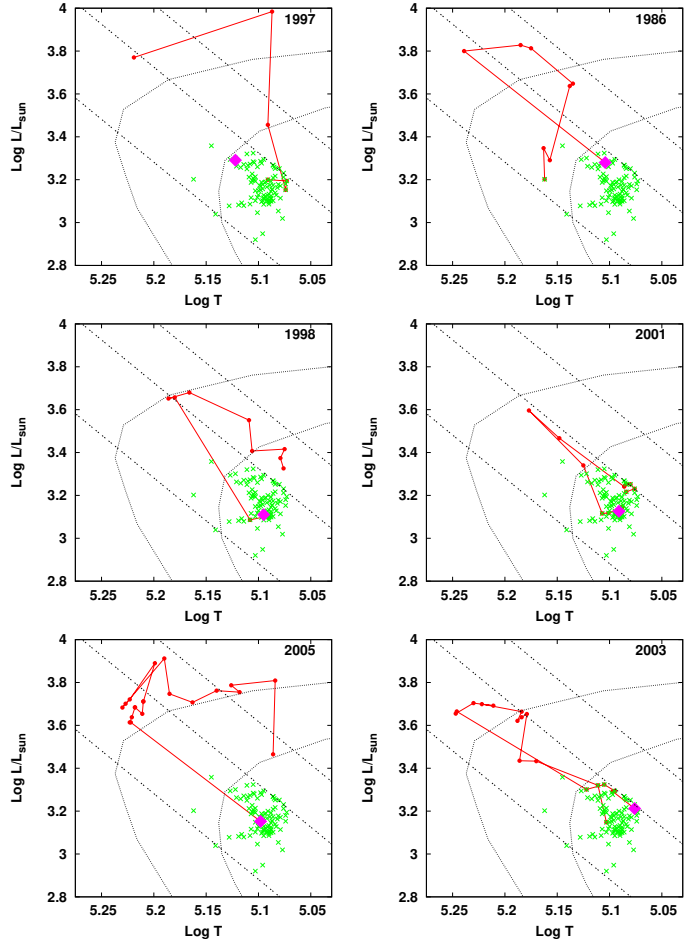}
   \caption{Six evolutionary tracks for the hot white dwarf in the symbiotic system AG Dra are presented. The crosses indicate the temperature and the luminosity of the white dwarf during the quiescence phase. The solid lines show the evolution during the cool outbursts (1997, 1998 and 2005) and hot outbursts (1986, 2001, 2003). The diamonds mark the beginning of the each outbursts. The dashed lines represent the constant radius tracks of 0.06, 0.10 and 0.14 Solar radii (from left to right). The evolutionary tracks (Vassiliadis et al. 1994) of white dwarfs with masses $M=0.593\,M_{\odot}$ and $M=0.640\,M_{\odot}$ are plotted as dotted lines.}
   \label{ltew}
\end{figure} 

Basing on Zanstra's method we have made a new attempt to compute the temperature of the hot ionizing source. Following \citet{osterbrock} we found the relation between the number of the photons emitted by the neutral hydrogen and the single ionized helium and the observed fluxes in H$\beta$ and HeII$\lambda4686$\AA\, emission lines. As a result we have obtained semi-empirical relations 

$$T\times10^{-4}= 4.83 + 2.04 \sqrt{65.04 \frac{F\left(HeII4686\right)}{F\left(H\beta\right)} - 0.08},$$
for the temperature range $50000 < T[K] < 140000$, and
$$T\times10^{-4}=12.41 \left(\frac{F\left(HeII4686\right)}{F\left(H\beta\right)}\right)^3 - 14.17 \left(\frac{F\left(HeII4686\right)}{F\left(H\beta\right)}\right)^2 +$$  
$$+ 23.04\left(\frac{F\left(HeII4686\right)}{F\left(H\beta\right)} \right) + 7.92$$ 
for $100000 < T[K] < 300000$. The fitted curves are shown on Figure~1.

Using the \textit{B-case} recombination equation we can compute the luminosity of the hot star separately from both $H\beta$ and HeII $\lambda4686$\AA\, and can plot the status of the hot star on the H-R diagram.

\section{Results}

The time evolution of the V magnitude of the system, as well as the temperature, the luminosity and the radius of the white dwarf is shown in Figure~2. The differences between the two types of outbursts are clearly visible in the \textit{U-band} photometry. The 1997, 1998 events are brighter than 2001, 2002 and 2003 ones. The duration time of 1997, 1998 and 2005 outbursts ($\sim200$ days) is significantly longer than the one of 2001, 2002 and 2003 outbursts ($\sim30$ days). The temperature reached by white dwarf during the 2002 and 2003 outbursts is higher than during the rest of outbursts, but the luminosities observed in 1997 and 2005 are the largest among the events shown here. The significant differences are visible in the hot star pseudo-photosphere radius. During the 2001-2003 outbursts the radius stays at the same level ($R_h=0.086\,R_{\odot}$) (marked with a dashed horizontal line), but it growth significantly up to $\sim0.22\,R_{\odot}$ in 1997, 1998 and 2005.

The evolution on the H-R diagram for the six different outbursts is graphicaly presented in Figure~3. The left column panels present the cool outbursts, the right column panel -- the hot ones. We have included the 1986 event (data taken from \citet{gr}), which is similar to 2002 and 2003 outbursts. Both types of outburst are characterised by an increase of the white dwarf luminosity and temperature during the first phase of the outburst. At the begining of the cool and hot outbursts the white dwarf evolves with the almost constant radius and increasing temperature and luminosity. After the star reaches maximum temperature and luminosity, two scenarios are possible for the further outburst evolution. According to the first one (hot outbursts), the cooling process fallows almost the same path to the quiet stage. According to the second scenario (cool outbursts) after the increase of the temperature and the luminosity a constant luminosity phase is observed.

Our analysis shows two kinds of outbursts -- cool and hot. As cool outbursts we can consider the 1997, 1998 and 2005 events, while the 2001, 2002, and 2003 outbursts belong to the hot ones. During the hot outbursts we do not observe the predicted by the theory of the thermonuclear outburst, constant luminosity phase of the white dwarf.



\acknowledgements 
This work is supported by the Polish MNiSW Grant N203 018 32/2338.


\end{document}